\documentclass{emulateapj}

\lefthead{Bekki \& Norris}
\righthead{Helium enrichment in $\omega$ Centauri}

\begin{document}

%\title{Searching for fossil records  of reionization by 
%wide-field imaging of galactic stellar halos}
\title{The origin of the double
main sequence in $\omega$ Centauri:  
Helium enrichment due to gas fueling
from its ancient host galaxy ?}

\author{Kenji Bekki} 
%\author{Kenji Bekki \altaffiltext{1}} 
\affil{
School of Physics, University of New South Wales, Sydney 2052, Australia}
\email{bekki@bat.phys.unsw.edu.au}

\and

\author{John E. Norris}
%\author{John E. Norris \altaffiltext{2}}
\affil{Research School of Astronomy \& Astrophysics,
        The Australian National University,
        Mt Stromlo Observatory, 
        Cotter Road, Weston Creek,   ACT 2611, Australia}
\email{jen@mso.anu.edu}

%\altaffiltext{1}{School of Physics, University of New South Wales,
%Sydney 2052, Australia, bekki@bat.phys.unsw.edu.au}
%\altaffiltext{2}{Research School of Astronomy \& Astrophysics,
%Australian National University,
%Cotter Road, Weston Creek,    ACT 2611, Australia, gdc@mso.anu.edu.au}

\begin{abstract}

 Recent observational studies of $\omega$ Centauri
by {\it Hubble Space Telescope} have discovered
a double main sequence in the color magnitude diagrams (CMDs)
of its stellar populations.
The stellar population with the blue main sequence (bMS)
is observationally suggested to have a helium abundance
much larger,
by $\Delta Y\sim 0.12$, than that of the red main sequence (rMS).
By using somewhat idealized models 
in which stars of the bMS are formed from gas ejected from
those of the rMS,
we quantitatively investigate whether the helium overabundance
of the bMS
can result from  self-enrichment from massive AGB stars,
from mass loss of very massive young stars, 
or from type II supernovae
within $\omega$ Cen.
We show that as long as the helium enrichment
is due to ejecta from the rMS formed earlier than the bMS,
none of the above three enrichment scenarios can explain
the observed properties of the bMS 
self-consistently for reasonable IMFs. 
The common,  serious problem in all cases 
is that the observed number fraction of the bMS can not be explained
without assuming unusually top-heavy IMFs. 
This failure of the self-enrichment scenarios
implies  that most of the helium-enriched
gas necessary for the formation of the bMS originated from
other external sources.
We thus suggest a new scenario that most of the second generation of
stars (i.e., the bMS) in $\omega$ Cen could be formed from
gas ejected from field stellar populations 
that surrounded $\omega$ Cen when it  was
a nucleus of an ancient dwarf galaxy.

\end{abstract}


\section{Introduction}

One of the most remarkable results of recent observational
studies of $\omega$ Cen is that it shows a double main sequence
(DMS) in the color magnitude diagrams (CMDs) of its stellar populations
(e.g., Anderson 1997; Bedin et al. 2004).
Bedin et al. (2004) proposed four possible scenarios
for the origin of the DMS:
(1) Theoretical isochrone models or data calibration are 
in error, (2) Stars on the bluer main sequence (bMS) of the DMS
are a super metal-poor ([Fe/H]$ \ll -2.0$) population,
(3) The bMS represents a very helium-rich ($Y\ge0.3$) population,
and (4) The bMS represents a background stellar population  about $1-2$ kpc
behind $\omega$ Cen.
A number of recent investigations have suggested that the above
scenario (3) (hereafter referred to as the ``helium pollution scenario''
and abbreviated to ``HEPS'' for convenience)
is the most  promising among the four
(e.g., Lee et al. 2005; Norris 2004; Piotto et al. 2005).

In the HEPS, there were two epochs of major star formation
in the history of $\omega$ Cen.
The first generation of stars  was  formed from proto-globular cluster (GC)
cloud(s)
to become stars on the red main sequence (rMS)
whereas the second generation of stars was
formed from gas ejected from the rMS.
One of the key questions related to  the HEPS is whether and how  
the observationally suggested large helium enrichment ($\Delta Y \sim 0.12$)
for the bMS can be obtained 
from the rMS with normal $Y$ ($ \sim 0.24$) 
in the star formation history of $\omega$ Cen
(e.g., Norris 2004). There are three  candidates
for the helium overabundance 
(Norris 2004; D'Antona et al. 2005; Piotto et al. 2005): 
(1) AGB stars with the masses larger than $6 {\rm M}_{\odot}$,
(2) stellar winds associated with massive stars during their early
evolutionary phases, 
and (3) type II supernovae (SNe II).
It is however unclear which of the above three is the most 
reasonable and realistic in the HEPS given a  lack of
extensive theoretical studies on the three candidates.

The purpose of this Letter is to investigate the three possibilities 
in a quantitative manner and thereby discuss which is the most promising
as the cause of the bMS (and the DMS) in the
context of the HEPS.
In this investigation,
(1) the possible $Y$ value 
observationally suggested for $\omega$ Cen (Norris 2004; Piotto et al. 2005) 
and (2) the observed  number fraction
of the bMS (Bedin et al. 2004)
are used to constrain the best possible
initial mass function (IMF) of forming stars 
in the HEPS. 
We do not intend to discuss extensively the observed abundance pattern 
of the bMS and the rMS in this paper, because chemical yield tables
for AGB stars with helium-rich ejecta ($Y > 0.35$) 
are not currently available.
We will discuss this point in our 
forthcoming papers (Bekki \& Norris 2005; BN).
Previous theoretical studies demonstrated that if GCs lose more than
50\% of their initial masses, they will  disintegrate
(e.g., Geyer \& Burkert 2001).
We also use this result as a constraint for globular cluster 
IMF in the HEPS.
We demonstrate that none of the above three candidates can 
explain the above constraint (2) without the modeled $\omega$ Cen 
disintegrating.

\section{The possible three polluters}

Firstly we summarize the possible three ``polluters'' which can significantly
enrich $\omega$ Cen with helium if the second  generation of stars 
(i.e., the bMS) was 
formed from their ejecta. In order for the observationally
suggested large value of $Y$ ($\sim 0.35$)  to be explained,
we need to select stellar objects whose ejecta have $Y$ significantly
higher than 0.24. 
Figure 1 shows the helium mass fraction ($Y$) 
in the ejecta of three candidates; (1) AGB stars
(van den Hoek \& Groenewegen 1997, hereafter  referred to as VG97);
Ventura \& D'Antona 2005,  as VA05),
(2) massive stars (Schaller et al. 1992, S92), 
and (3) SNe II  (Woosley \& Weaver 1995, WW95).
Based on the values listed in these papers,
$Y$ is derived for each stellar mass.
The theoretical predictions of $Y$ for AGB stars is at most
0.28 for VG97 
and 0.32 for VA05.
D'Antona et al. (2005) however suggested that more massive AGB
stars with the masses of $6-7 {\rm M}_{\odot}$ can have
$Y=0.40$ (see also Lattanzio et al. 2004).
 We therefore consider that  massive AGB stars
can be the polluters in  $\omega$ Cen but that 
lower mass AGB objects  ($\le 5 {\rm M}_{\odot}$) with
smaller $Y$ can not.

The observed difference in metallicity ([Fe/H])
between the rMS and the bMS (Piotto et al. 2005) suggests that
the above  scenario (1) alone can not simply explain
the metallicity difference, because metallicity of AGB ejecta
do not differ from those of their host stars. 
Therefore, full chemical evolution models
including the contribution of the three polluters
(and mixing between the ejecta of the polluters and
fresh gas) need to be 
considered if we try to explain fully self-consistently both 
the helium 
and metallicity one between the rMS and the bMS.
We however discuss advantages and disadvantages of
each of the three polluters in the helium enrichment process separately.
This will better permit  
us to disentangle the contribution of the different polluters in
producing He-enhanced stars and thereby to analyze advantages and
disadvantages of each scenario in explaining the helium abundance
of the bMS. The entire chemical evolution history for different
elements (e.g., ${}^{12}$C) in $\omega$  Cen will be discussed in
our future papers (BN).

\subsection{AGB stars}

The first question is whether the number fraction of 
the second  generation of stars (i.e., the bMS fraction)
can be as large as $0.25$ (Bedin et al. 2004)
for a reasonable IMF without $\omega$ Cen
being disintegrated following  mass loss from massive stars and SNe II. 
In order to estimate the  mass fraction ($f_{\rm AGB}$)
of AGB progenitor stars with
masses ranging from $6 {\rm M}_{\odot}$
to  $7 {\rm M}_{\odot}$ in a GC with the total mass of $M_{\rm cl}$,
we assume an 
IMF in number defined
as $\psi (m_{\rm I}) = A{m_{\rm I}}^{-s}$,
where $m_{\rm I}$ is the initial mass of
each individual star and the slope $s=2.35$ corresponds to the Salpeter IMF.
The normalization factor $A$ is a function of $M_{\rm cl}$,
$m_{\rm l}$ (lower mass cut-off), and $m_{\rm u}$ (upper one).
A is expressed as
$A=\frac{M_{\rm cl} \times (2-s)}{{m_{\rm u}}^{2-s}-{m_{\rm l}}^{2-s}}$,
where $m_{\rm l}$ and $m_{\rm u}$ are  set to be   $0.1 {\rm M}_{\odot}$
and  $120 {\rm M}_{\odot}$, respectively.
Although the number fraction of low-mass stars
on the bMS can depend on the forms of IMFs, we adopt the above IMF
and show the results.
This is mainly because we can show more clearly the roles of
IMF slopes in controlling the mass fraction of AGB stars
for the adopted IMF in this preliminary
study.
The models with the IMF proposed by Kroupa et al. (1993)
show a rather small  fraction ($\sim 0.1$ \%) of AGB stars in comparsion
with the models with the above IMF. 

In this AGB pollution scenario, 
we assume that 
(1) the gas ejected from AGB stars in the first generation
(i.e., the rMS) is {\rm all} used (i.e.,
a star formation efficiency of 100 \%) for the formation
of the second generation (i.e., the bMS)
and (2) the total amount of gas ejected from a AGB star
is the same as the stellar mass.
This second assumption is reasonable because more than 80 \%
of the stellar mass is  ejected as AGB winds for AGB stars
with  $m_{\rm I} > 6 {\rm M}_{\odot}$  
(e.g., VG97).
We also assume two different  slopes of IMF ($s$) for
the first ($s_{1}$) and the second ($s_{2}$) generations of stars
in order to 
investigate the maximum value of 
$f_{\rm 2nd}/(f_{\rm 1st}+f_{\rm 2nd})$,
where $f_{\rm 1st}$ and $f_{\rm 2nd}$ are the number of stars (per unit mass)
with masses smaller than $0.88 {\rm M}_{\odot}$
(corresponding to stars older than $\sim 12$ Gyr)
for the first generation and for the second one, respectively.

Figure 2 shows that the smaller the IMF slope $s_{1}$ is 
(i.e., the IMF becomes more ``top-heavy''),  
the larger the number fraction of the second generation 
($f_{\rm 2nd}/(f_{\rm 1st}+f_{\rm 2nd})$) becomes
for a  fixed $s_{2}$ of $2.35$.
The ratio of the final cluster mass to its initial mass ($f_{\rm rem}$)
however also becomes smaller for  smaller $s_{1}$.
As a result of this,  $f_{\rm rem}$ 
of the models with $s_{1} \le  1.75$ become smaller than
the threshold value (0.5) below which star clusters can be disintegrated
as the result of  mass loss 
(e.g., Geyer \& Burkert 2001): No models can be located
within the regions with $0.25 \le f_{\rm 2nd}/(f_{\rm 1st}+f_{\rm 2nd}) \le 0.35$
and with $f_{\rm rem} \ge 0.5$ where
$\omega$ Cen can survive possible disintegration and have
the bMS fraction consistent with observations.
The number fraction of the second generation can not be larger than
0.16, even for  $s_{2}$ = $5.35$ (i.e., extremely bottom-heavy)
and $s_{1}$ = $1.95$ for which  $\omega$ Cen can manage to
survive  disintegration due to mass loss.
Thus the results in Figure 2 suggest that the observed fraction
of the bMS can not be explained  simply by the AGB pollution scenario
for reasonable IMFs.

If $m_{\rm u}$ is as low as $8 {\rm M}_{\odot}$,
$\omega$ Cen can not disintegrate owing to mass loss from SNe II
for  any values of $s_{1}$ and $s_{2}$.
It is found that $f_{\rm 2nd}/(f_{\rm 1st}+f_{\rm 2nd})$ is
about 0.29 for the model with $m_{\rm u} =8 {\rm M}_{\odot}$,
$s_{1}=1.35$, and $s_{2}=2.35$.
This result suggests that if $\omega$ Cen was formed
with a very unusual  IMF (i.e., with no or little SNe II),
the observed bMS fraction can be reproduced without
disintegration of $\omega$ Cen.
The above  model with no or little  stellar remnants
(black holes and neutron stars) of massive stars,
however, would not be   consistent with the presence of
neutron stars in $\omega$ Cen (e.g., Haggard et al. 2004).

\subsection{Massive  stars}

The helium mass fraction in the stellar winds of massive
stars with  $m_{\rm I} > 85  {\rm M}_{\odot}$ can become
as high as 0.4 with a  maximum of 0.49 for  $m_{\rm I} = 120 {\rm M}_{\odot}$
(Schaller et al. 1992). Although  there is no problem with $Y$ in
this scenario, 
the mass fraction of massive stars
with $m_{\rm I} > 85  {\rm M}_{\odot}$ can not be as large as 
$\sim 0.25$ for reasonable IMFs ($1.95 \le s \le 2.35$) 
for which $\omega$ Cen does not
disintegrate. 
Figure 3 shows that the mass fraction of the ejecta of the massive
stars is at most $0.2$ for a  plausible range of IMF slopes.
Given the fact that the observed star formation efficiency 
in the Galaxy can be at most   
$\sim 0.4$ (Wilking \& Lada 1983 for the $\rho$ Oph cloud),
this result suggests that the observed fraction of the bMS ($\sim 0.25$)
is unlikely
to be explained by this scenario. 

A more  serious problem
for this stellar wind scenario 
is that there would be no or little difference 
in the mean metallicity
between the first generation of stars and the ejecta of these objects
(Schaller et al. 1992): Piotto et al. (2005) demonstrated that 
there is a metallicity difference by $\Delta$[M/H]$\sim 0.3$
between the rMS and the (more metal-rich) bMS in $\omega$ Cen. 
Another possible problem of this stellar wind scenario
is that  the ejecta should  be very quickly
converted into stars (within less than $\sim 10^7$ yr)
between the stellar wind phases of very massive stars and 
the onset of SNe II that can blow away the ejecta from
$\omega$ Cen.

\subsection{Type II supernovae }

The helium mass fraction ($Y$) in the ejecta of SNe II 
stars with  $12 \le m_{\rm I}/{\rm M}_{\odot} \le 40$ 
for metal-poor stars with $Z=0.01 Z_{\odot}$ 
ranges from 0.37 to 0.45.
(WW95). Although this scenario has no problems with $Y$,
the abundance by mass for  heavier elements in the ejecta 
is too high to be consistent with observations.
${}^{12}$C and ${}^{40}$Ca in WW95 are as high as $\sim 0.01$ and $0.001$,
respectively, which are  higher than the solar values
($\sim 3.0 \times 10^{-2}$ and $\sim 6.0 \times 10^{-5}$,
respectively; WW95) and thus much higher than those of
$\omega$ Cen. 
This inconsistency with the observed metallicities
suggests that the bMS can not be formed directly from
the ejecta of SNe II of the rMS. 
If the ejecta of SNe II  can be mixed with the primordial gas
used for the formation of the rMS, the abundance of
${}^{12}$C and ${}^{40}$Ca of the bMS formed
from the mixed gas can be significantly reduced.
However, the helium abundance would  also be significantly
reduced in the process of mixing. Therefore it can be
concluded that this scenario is highly unlikely
to explain the observed abundance of the bMS.
Furthermore this scenario has  difficulty in explaining
the observed fraction of the bMS without assuming
top-heavy IMFs, just as the other two scenarios do.

\placefigure{fig-1}

\section{Discussion: Alternative scenarios}

\subsection{External pollution in the nucleus of a dwarf galaxy}

The present study shows that the proposed three scenarios 
all fail to explain self-consistently the observed
properties of the bMS in $\omega$ Cen.
The AGB scenario appears to be  the most   promising 
among the three, 
though it remains unclear whether
this scenario can explain the observed abundances 
other than helium in $\omega$ Cen.
The most  serious 
problem of the AGB scenario 
is that the total mass  of the ejecta of more massive AGB
stars from the rMS is too small to be consistent with the observed
fraction of the bMS of $\omega$ Cen. If however  we relax
the adopted assumption that {\it all stars of the bMS originate
from AGB ejecta of the rMS initially within  $\omega$ Cen}, the problem
of the AGB scenario can be significantly alleviated.
The problem of the bMS fraction can be solved if
the original total mass of the rMS was  
larger (by a factor of $\sim 10$)
than the present mass (i.e., if $\omega$ Cen was a super-giant
GC)
and if most of the stars of the rMS  were removed 
from $\omega$ Cen 
after their ejecta  were used for the formation
of the bMS.

We here discuss a scenario in which {\it $\omega$ Cen was
the nucleus of a nucleated dwarf galaxy (dE,N) where 
the AGB ejecta of the central field stellar populations surrounding the  
(proto-) $\omega$ Cen  were  consumed for star formation of the bMS.}
In this scenario, the ancient host dE,N of $\omega$ Cen 
has been   completely destroyed by the Galactic tidal field
so that only its nucleus (i.e., $\omega$ Cen) is now  observed
(e.g., Bekki \& Freeman 2003).
Not only the ejecta of the rMS but also those from stellar populations
surrounding the nucleus can be used for the formation
of the bMS: the above bMS fraction problem
should not be  so serious  in this scenario.
The bMS can  be more metal-rich than the rMS if 
the mean abundance of the field stellar 
populations and gas in the dwarf is slightly higher than that of the rMS.

The key question in this  scenario 
is how much gas ejected from  the central field
stars can be converted into the bMS without being removed by
energetic outflow of AGB stars in the central regions of 
$\omega$ Cen's host dE,N. Recent numerical simulations on
transformation  from a dE,N into $\omega$ Cen have suggested that
$\omega$ Cen's host dE,N has $M_{\rm B} \sim -14$ mag,
corresponding to the total stellar mass 
of  1.25$\times$ $10^8$ M$_{\odot}$ (e.g., Bekki \& Freeman 1993).
If the total mass of the bMS is about 30\% of
present-day $\omega$ Cen's mass
(= 5.0 $\times$ $10^6$  M$_{\odot}$; Meylan  et al. 1995),
it is necessary for only 1.2\% of the stellar mass of the dE,N
to have been  converted into the bMS.
Since the mass fraction of 
more massive AGB stars in the dE,N ($> 0.02$)  for a 
reasonable IMF  
can be significantly larger than the above value ($\sim 0.01$),
the observed bMS fraction is not a problem in this  new scenario.

The central escape velocity ($V_{\rm esc}$) 
of this host embedded in a massive dark
matter halo with the total mass of $10^8-10^9  M_{\odot}$ is $50-90$ km s$^{-1}$
depending on the  inner profile of dark matter halo (Bekki 2005).
The derived $V_{\rm esc}$  is 
significantly larger than the observed maximum wind velocity
of $\sim 40$ km s$^{-1}$ for AGB stars (e.g., Loup et al. 1993).
Therefore, the AGB ejecta of the dE,N is highly likely to be 
trapped within the central region of the dE,N and consequently
converted into new stars (i.e., the bMS). 
Although extensive numerical simulations on the central gas dynamics
of a dE,N are necessary to confirm that gas ejected from field stars
of the dE,N 
can be transferred into the nuclear region and converted into new stars there,
we propose that the bMS can originate from gas ejected from  
stars within  
the ancient dE,N whose nucleus was  $\omega$ Cen.

\subsection{Helium sedimentation}

The local helium abundance 
of the remaining gas of forming $\omega$ Cen might have been
significantly higher (leading to a bMS with higher helium abundance), 
if helium sedimentation due to
gravitational diffusion (which was originally proposed by
Fabian \& Pringle (1977) for a mechanism responsible for
gaseous abundance gradients in clusters of galaxies)
occurred  in $\omega$ Cen.
Chuzhoy\& Loeb  (2004) showed that gravitational diffusion
in the interstellar  medium of giant elliptical galaxies 
can result in the increase of helium abundance  
by a factor of $1+{0.2(T/10^7)}^{1.5}/F_{\rm B}$,
where $T$ and $F_{\rm B}$ are the gas temperature
and the suppression factor (likely to be $>5$)
of the gravitational diffusion by the  magnetic field. 
These results suggest that a significant  increase of helium
abundance is highly unlikely within  GCs, which have
the central velocity dispersions of order of 10 km s$^{-1}$,
corresponding to $T \sim 10000$ K.

\section{Conclusion}

We have shown that the observed bMS fraction 
of $\omega$ Cen can not be  simply explained by 
any HEPS in which the bMS was formed from ejecta of the rMS.
We accordingly have suggested an ``external pollution'' scenario
in which the bMS was formed from gas that was initially within
the central region of $\omega$ Cen's host galaxy.
The question yet to be answered in this scenario is how
star formation could  proceed within  the rMS
when gas was transferred to the central region.
It would be possible that the bMS was formed 
outside yet close to the rMS 
as a star cluster and then merged with the rMS
in the central region of $\omega$ Cen's host galaxy.

\acknowledgments
We are  grateful to the anonymous referee for valuable comments,
which contribute to improve the present paper.
K.B. and J. E. N. acknowledge the financial support of the Australian Research 
Council throughout the course of this work.

\clearpage

%%%%%%%%%%%%%%%%%%%%%%% Figure Captions

\newpage
\plotone{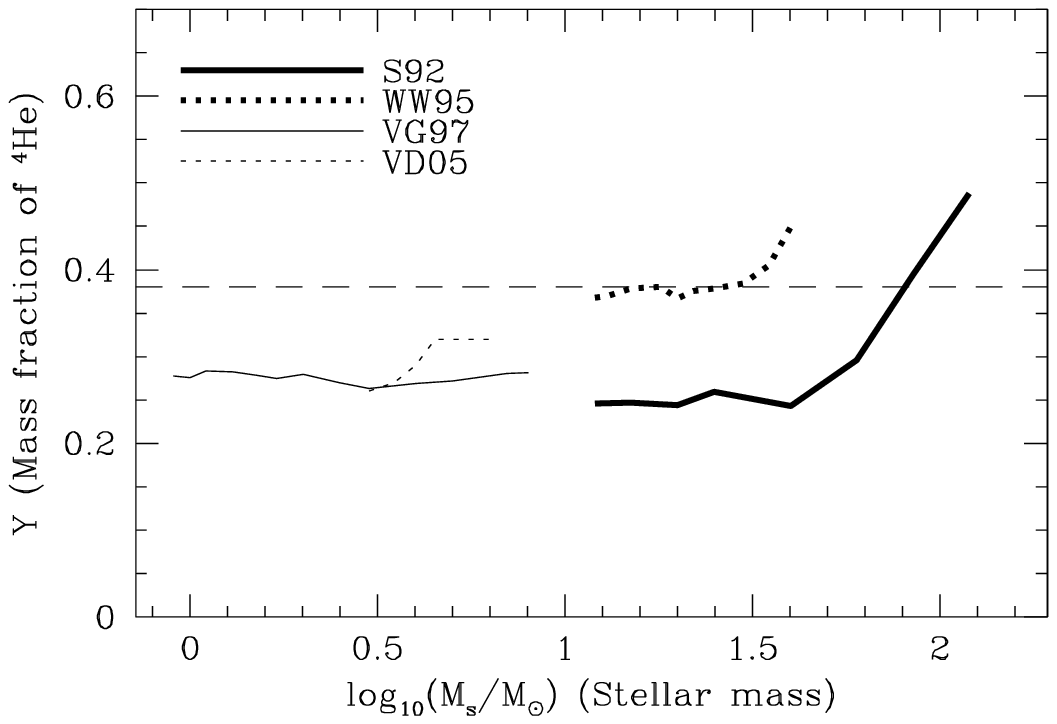}
\figcaption{
The mass fraction of helium $^{4}$He ($Y$) in the ejecta
of stellar winds of massive stars (thick solid), 
SNe II (thick dotted), and AGB stars (thin solid and dotted)
for different masses ($M_{\rm s}$).
For comparison, the observationally suggested
value of $Y$  for the bMS of 
$\omega$ Cen is shown by a dashed line.
$Y$ values were  calculated by using the tables
by Schaller et al. (1992) for massive stars,
Woosley \& Weaver (1995) for SNe II,
and  van den Hoek \& Groenewegen (1997)
(thin solid line)
and  Ventura \& D'Antona (2005)
(thin dotted line) for AGB stars. 
\label{fig-1}}

\newpage
\plotone{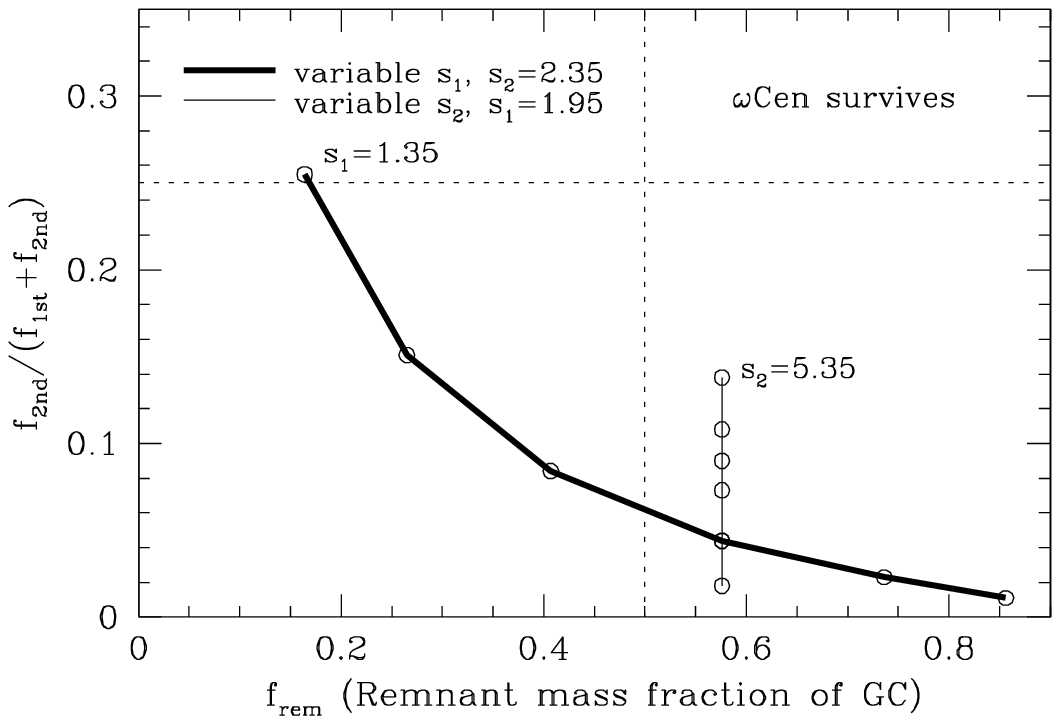}
\figcaption{
The locations of $\omega$ Cen   models with different IMFs 
in the $f_{\rm rem}$-$f_{\rm 2nd}/(f_{\rm 1st}+f_{\rm 2nd})$ plane.
Here  $f_{\rm rem}$, $f_{\rm 1st}$, and  $f_{\rm 2nd}$
represent the mass fraction of remnant stars
(i.e., low-mass, long-lived stars) after mass loss due to
supernovae, the number of stars of the first generation,
and that of the second. The vertical dotted line represents
the  boundary ($f_{\rm rem}=0.5$) beyond which 
$\omega$ Cen  can survive   
disintegration due to mass loss.
The horizontal dotted line represents the lower limit of the
observed bMS fraction of $\omega$ Cen in Bedin et al. (2004).
The upper right region surrounded by these two lines
is where any $\omega$ Cen model should lie to explain
observations (i.e., self-bounded and larger bMS fraction).
The models with $s_{1}$ = $1.35$ (left), 
$1.55$ (second from left), $1.75$ (third from left),
$1.95$ (third from right),
$2.15$ (second from right), and $2.35$ (right)
for a fixed $s_{2}$ of 2.35 (the Salpeter IMF)
are shown by a thick solid line.
The models with $s_{2}$ = $5.35$ (top), 
$4.35$ (second from top), $3.35$ (third from top),
$2.85$ (third from bottom),
$2.35$ (second from bottom), and $1.95$ (bottom)
for a fixed $s_{1}$ of 1.95 
are shown by a thin  solid line.
$s_{1}$ ($s_{2}$) is the slope
of the IMF in the form of  $\psi (m_{\rm I}) = A{m_{\rm I}}^{-s}$
for the first (second) generation of stars.
Note that all models have serious difficulties  in
reproducing both $f_{\rm rem} \ge 0.5$ and  
$f_{\rm 2nd}/(f_{\rm 1st}+f_{\rm 2nd}) \ge 0.25$
simultaneously.
\label{fig-2}}

\newpage
\plotone{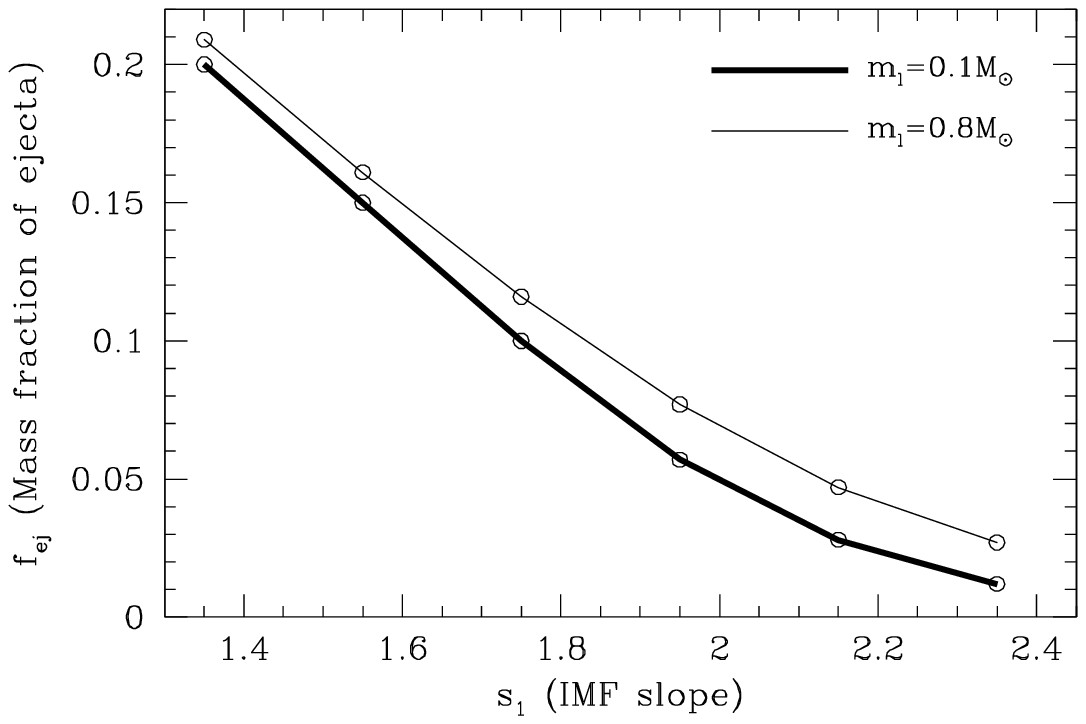}
\figcaption{
Dependences of the mass fraction of ejecta from massive stars
with $85 \le  m_{\rm I}/{\rm M}_{\odot} \le 120$
on the IMF slope ($s_{1}$) of the first generation of stars
for the models with $m_{\rm l} = 0.1 {\rm M}_{\odot}$ (thick)
and $m_{\rm l} = 0.8 {\rm M}_{\odot}$ (thin).
\label{fig-3}}

\end{document}